\documentclass[a4paper]{jpconf}
\usepackage{graphicx}
\usepackage{bm}

\newcommand{\Tc}{T_{\mbox{\scriptsize C}}}
\newcommand{\TN}{T_{\mbox{\scriptsize N}}}

\newcommand{\hac}{h_{\mbox{\scriptsize ac}}}

\begin{document}
\title{Anomalous slow dynamics in the metallic helimagnet Gd$_{1-x}$Y$_{x}$}

\author{T Yamazaki$^1$, Y Tabata$^2$, T Waki$^2$, H Nakamura$^2$}

\address{$^1$Institute for Solid State Physics, University of Tokyo, Chiba 277-8581, Japan}
\address{$^2$Dept. of Materials Science and Engineering, Kyoto University, Kyoto 606-8501, Japan}

\ead{t.yamazaki@issp.u-tokyo.ac.jp}

\begin{abstract}
AC-suseptibility measurements were performed in the metallic helimagnet Gd$_{1-x}$Y$_{x}$ alloy.
A remarkable increase of the imaginary part of the ac-suseptibility was observed in the temperature range of the helimagnetic phase.
Moreover, a strong nonlinearity of the magnetization was observed at paramagnetic-helimagnetic transition temperature $\TN$.
On the other hand, These anomalous behavior were not observed in similar rare-earth helimagnets Ho and Ho$_{1-x}$Y$_{x}$.
It strongly suggests that the weak magnetic anisotropy of the Gd-moments is responsible for the anomalous slow dynamics in the helimagnetic phase and for the strong nonlinearity of the magnetization at $\TN$.
The slow dynamics may result from a chiral-domain motion, or a long-time variation of the period of the helimagnetic structure.

\end{abstract}

\section{Introduction}
The rare-earth alloy Gd$_{1-x}$Y$_{x}$ exhibits interesting magnetic properties, such as the Lifshits transition between the long-period helical magnetic state and the ferromagnetic state in zero magnetic field, and has been investigated extensively for decades \cite{ Legvold1980, Legvold1980a, Bates1985,Melville1992, Andrianov2008}.
The crystal structure of the Gd$_{1-x}$Y$_{x}$ alloys is hexagonal-closed-pack (hcp) one.
Many heavy rare-earth hcp metals and their yttrium alloys R$_{1-x}$Y$_x$ exhibit long-period modulated structures,which can be explained by the geometry of their Fermi surfaces \cite{Fretwell1999, Hughes2007}.
A phase transition from a modulated magnetic structure to a collinear ferromagnetic one is generally found in the heavy rare-earth metals.
Most of these transitions are actually first order phase transitions accompanied by lattice distortions because of the strong spin-orbit interaction \cite{Evenson1969}, which cause the strong magnetic anisotropy.
Among of them, Gd$_{1-x}$Y$_{x}$ alloys are important exceptions, 
 where the spin-orbit interaction does not act because Gd$^{3+}$-ion has a zero orbital-angular moment ($L=0$) \cite{Jensen}.
And hence, a continuous transition from the modulated structure to the collinear ferromagnetic one can be realized in Gd$_{1-x}$Y$_{x}$.
An interesting second-and-a-half order transition, which may be a Lifshitz transition, was reported actually in Gd$_{1-x}$Y$_{x}$ \cite{Andrianov2008}.
In the Y-concentration range of 0.32 $< x <$ 0.40, Gd$_{1-x}$Y$_{x}$ exhibits a magnetic transition at $\TN$ from the paramagnetic state to the proper-type helimagnetic state and a successive transition at $\Tc$ to the collinear ferromagnetic state.
The propagation vector $\bm {q}$ parallel to the c$^*$-direction continuously decreases to zero with decreasing temperature from $\TN$ to $\Tc$ \cite{Bates1985}.

In this paper, we report an anomalous slow dynamics and a strong nonlinearity of magnetization newly discovered in Gd$_{1-x}$Y${_x}$.
These anomalous phenomena most likely result from the very weak magnetic anisotropy of Gd-moments 
because they were not observed in the similar rare-earth helimagnets Ho and Ho$_{1-x}$Y${_x}$.
We speculate two possible origins of the anomalous slow dynamics in the Gd$_{1-x}$Y${_x}$.
The first one is a collective spin dynamics with keeping its helical spin arrangements,
which is a motion of the chiral-domains themselves.
The second is a long-time variation of the period of the helimagnetic structure.

\section{Experimental}
Single crystalline samples of Gd$_{0.62}$Y$_{0.38}$ were grown by the Czochralski pulling method with tetra-arc furnace.
The obtained crystals wrapped in the Ta films and sealed in the evacuated quartz tubes and annealed for 1 week at 973 K, and then, quenched in cold water.
Polycrystalline samples of Ho and Ho$_{0.60}$Y$_{0.40}$ prepared by arc melting with the mono-arc furnace were also used for the comparison study.
The size of Gd$_{0.62}$Y$_{0.38}$ samples for the ac-susceptibility measurements along the a$^*$- and c$^*$-axes are 0.9 $\times$ 0.3 $\times$ 1.9 mm$^3$  and 0.4$\times$0.2$\times$1.9 mm$^3$ respectively, and the size of Ho and Ho$_{0.60}$Y$_{0.40}$ samples are 0.5 $\times$ 0.5 $\times$ 9.0 mm$^3$.

The ac-susceptibility measurements were performed by using the SQUID-magnetometer (Quantum Design, MPMS) equipped in the Research Center for Low Temperature and Material Sciences in Kyoto University.
In the measurements, dc-magnetic-field was also applied parallel to the ac-magnetic-field.

\section{Results and Discussion}
\begin{figure}[b]
\begin{minipage}{18pc}
\includegraphics[width=18pc]{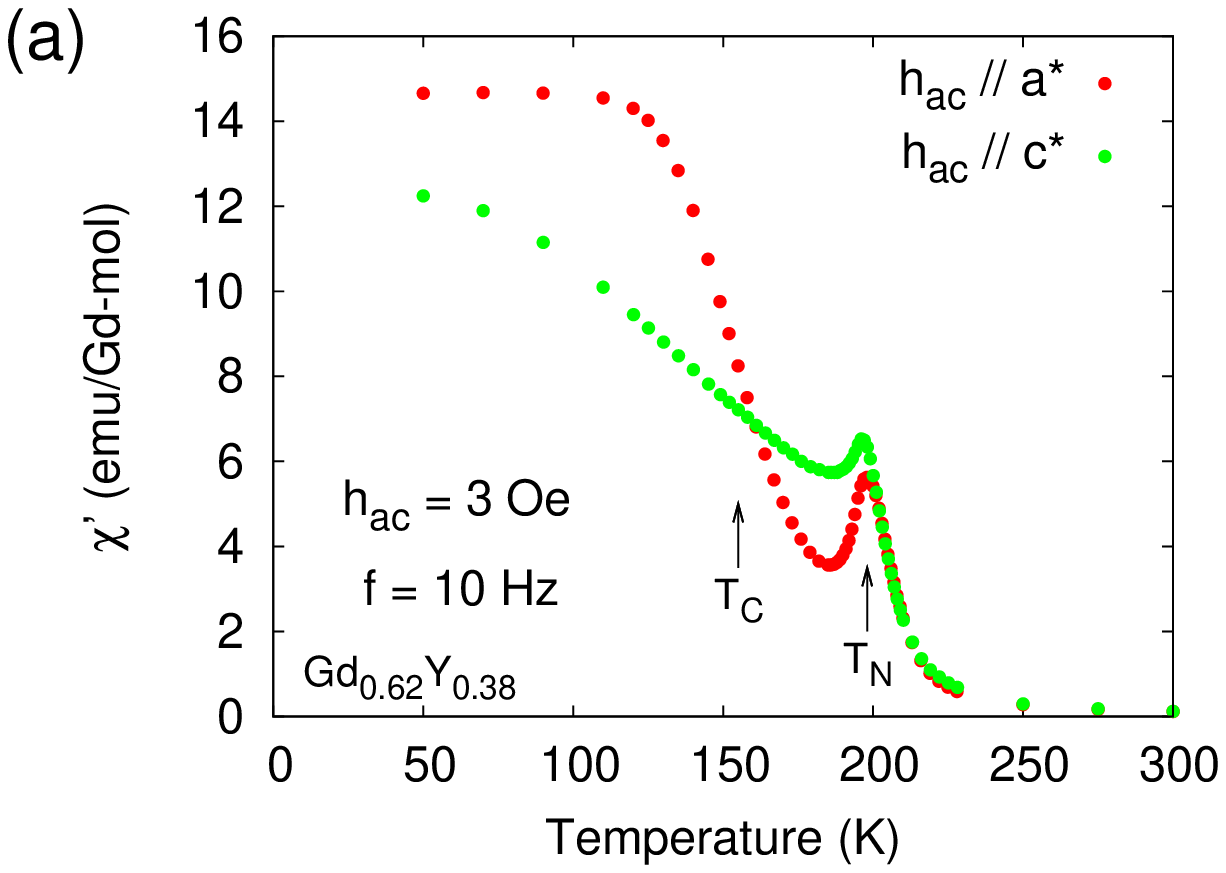}
\end{minipage}\hspace{2pc}%
\begin{minipage}{18pc}
\includegraphics[width=18pc]{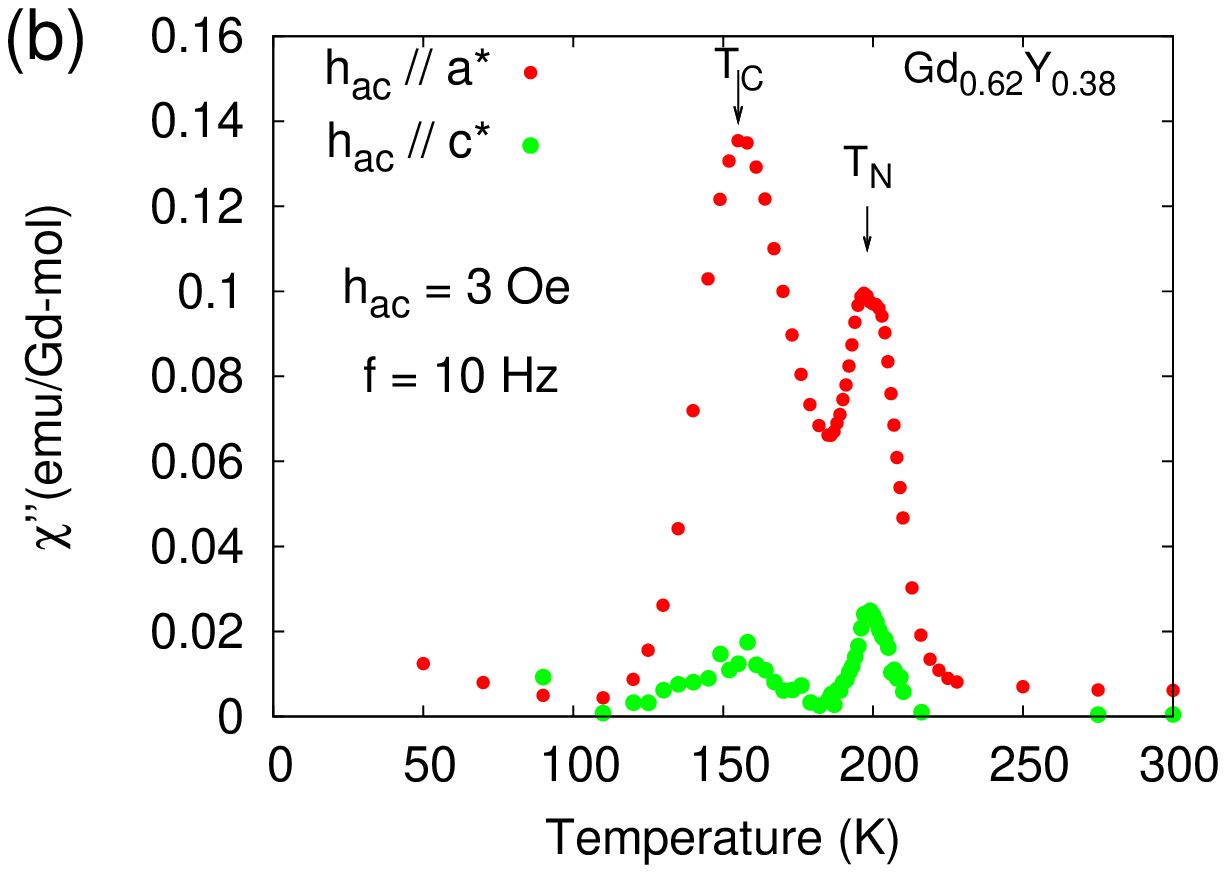}
\end{minipage} 
\caption{\label{AvsC}Temperature dependences of the real and imaginary parts of the ac-susceptibility, $\chi'$ and $\chi''$, in Gd$_{0.62}$Y$_{0.38}$ alloy with applying the ac-magnetic-field along the a$^{\ast}$- and c$^{\ast}$-directions.}
\end{figure}

Figures \ref{AvsC} (a) and (b) show temperature ($T$) dependences of the real and imaginary parts of the ac susceptibility, $\chi'$ and $\chi''$, in Gd$_{0.62}$Y$_{0.38}$ 
, respectively.
In the figures, we show the data with applying the ac-magnetic field along the a$^{\ast}$- and $c^{\ast}$-directions, being in and perpendicular to the helical plane.
In these measurements, the magnitude and frequency of the ac-magnetic-field $\hac$ and $f$ were 3 Oe and 10 Hz, respectively. 
In the $T$-dependences of $\chi'$, a sharp cusp-type anomaly is found at 198 K, corresponding to the para-helimagnetic phase transition at $\TN$.
A signature of the heli-ferromagnteic transition is found in $\chi'$ along the a$^{\ast}$-direction,
 a rapid increase of $\chi'$, around 155 K.
On the other hand, the increase of $\chi'$ along the c$^{\ast}$-direction is more gentle.
These are consistent with the fact previously reported, the ferromagnetic ordered moment lies around the c$^{\ast}$-plane.
The most striking feature we found in Fig. \ref{AvsC} is a remarkable increase of $\chi''$ in the whole $T$-range of the helimagnetic phase.
It indicates a presence of a slow dynamics in the helimagnetic phase.
It should be noted that the increase of $\chi''$ in the helical plane is much stronger than that perpendicular to the helical plane.

\begin{figure}[t]
\begin{center}
\includegraphics[width=24pc]{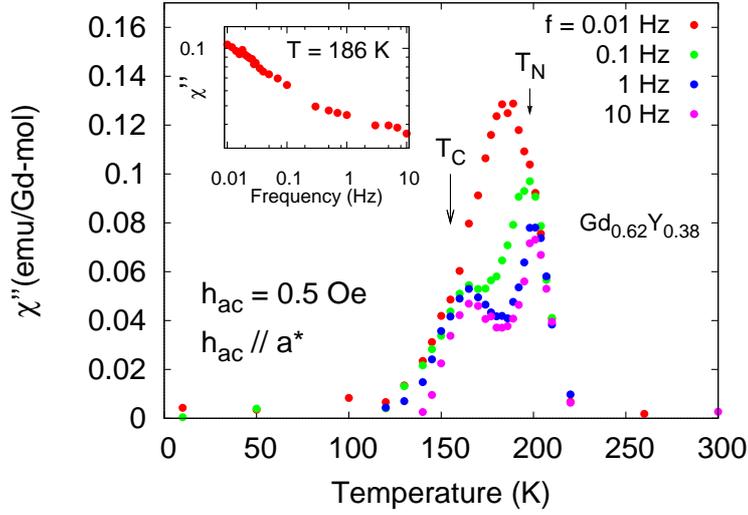}
\caption{\label{freqency}Temperature dependences of $\chi'$ of Gd$_{0.62}$Y$_{0.38}$ along the a$^{\ast}$-direction, being in-plane direction, with various frequency. Inset: Frequency dependence of $\chi'$ at 186 K with a double logarithmic scale.}
\end{center}
\end{figure}

To obtain further information of the anomalous slow dynamics, we performed ac-susecptibility measurements along the a*-direction with various frequency in the range of 0.01 Hz $\le f \le$ 10 Hz, shown in Fig. \ref{freqency}.
In this measurements, the smaller ac-magnetic-field of 0.5 Oe was applied to avoid the nonlinear effect appeared in the helimagnetic phase, which will be described later.
The increase of $\chi''$ becomes more remarkable  with decreasing the frequency from 10 Hz to 0.01 Hz.
At 0.01 Hz, a large broad peak of $\chi''$ is found around 186 K.
The results clearly indicate that the dynamics in the helimagnetic phase is extremely slow.
The log-log plot of the frequency dependence of $\chi''$ at 186 K is shown in the inset of Fig. \ref{freqency}.
$\chi''$ rapidly increases with decreasing the frequency down to 0.01 Hz, indicating that the characteristic frequency of the helimagnetic state is smaller than 0.01 Hz.

\begin{figure}[t]
\begin{minipage}{9pc}
\includegraphics[width=12pc]{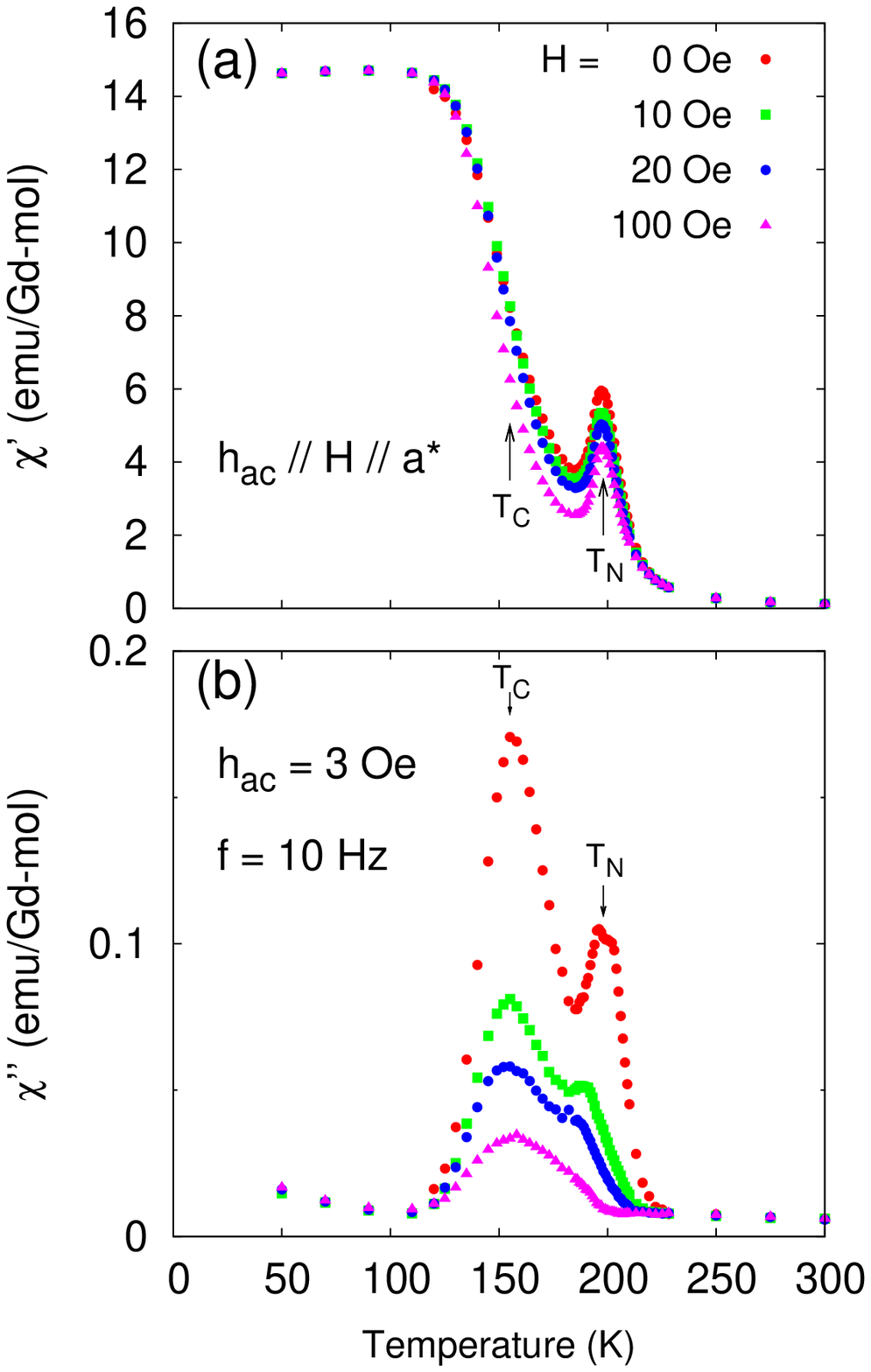}
\end{minipage}\hspace{2pc}%
\begin{minipage}{12pc}
\includegraphics[width=12pc]{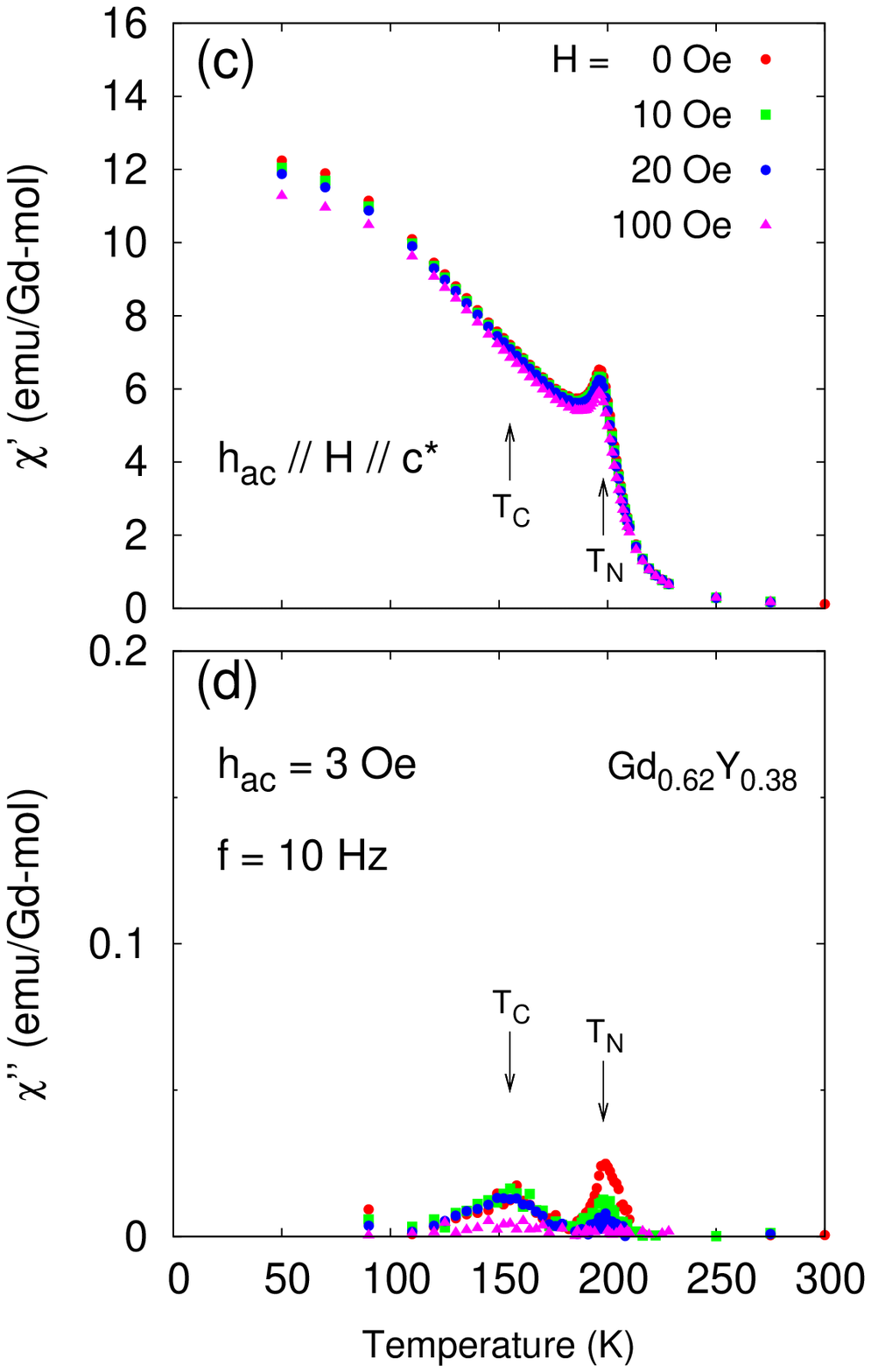}
\end{minipage} 
\begin{minipage}{14pc}
\includegraphics[width=14pc]{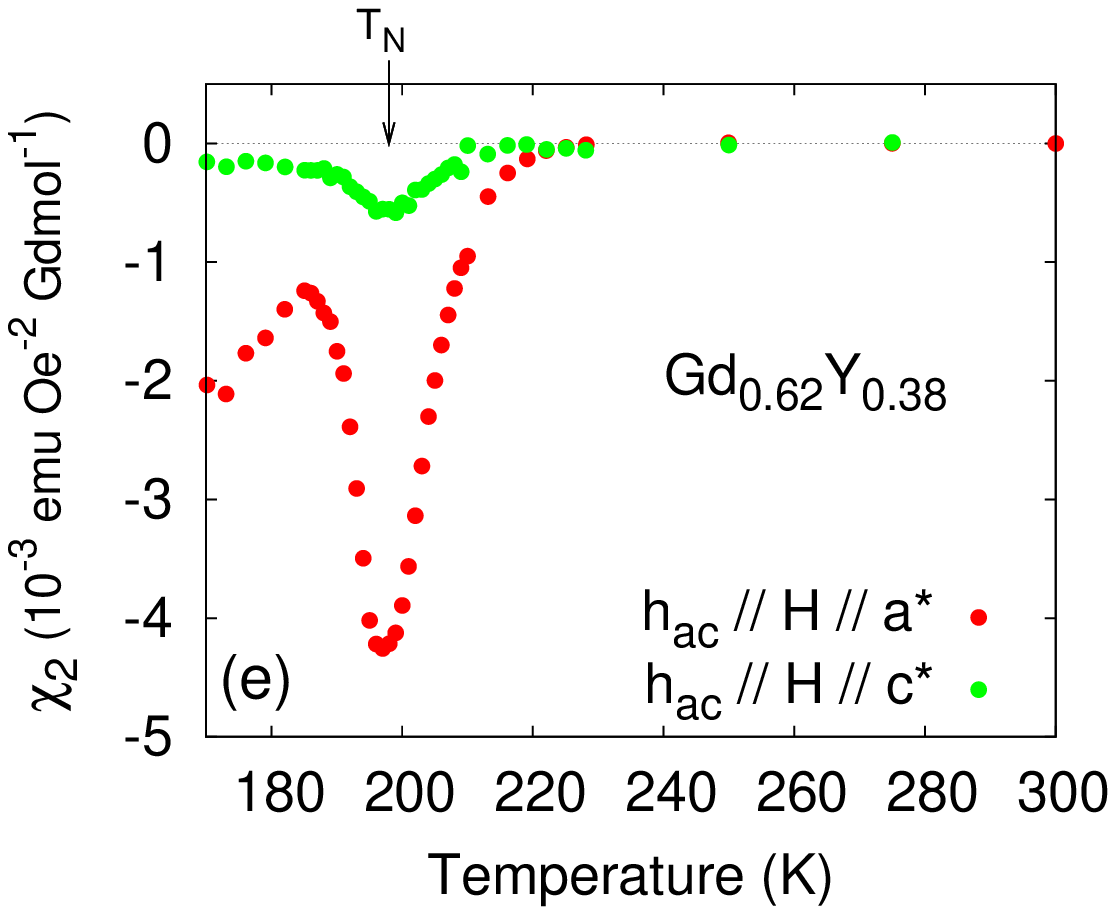}
\end{minipage}\hspace{2pc}%
\caption{\label{H-dep}(a)-(d) Temperature dependences of $\chi'$ and $\chi''$ of Gd$_{0.62}$Y$_{0.38}$ with applying the various dc-magnetic-field along the a$^{\ast}$-direction and c$^{\ast}$-direction, respectively. (e) The temperature dependences of the nonlinear susceptibility $\chi_2$ along the a*- and c*-directions.}
\end{figure}

Figure \ref{H-dep} (a)-(d) show temperature dependences of $\chi'$ and $\chi''$ of Gd$_{0.62}$Y$_{0.38}$ along the a$^{\ast}$ and c$^{\ast}$-directions, respectively, with applying the dc-magnetic-field of 0 Oe $\leq H \leq$ 100 Oe.
The magnitude and frequency of the ac-magnetic-field were 3 Oe and 10 Hz, respectively. 
Along the a$^*$-direction, $\chi'$ and $\chi''$ are strongly suppressed by the dc-magnetic-field in the helimagnetic state (see Fig.\ref{H-dep} (a), (b)).
The suppression of $\chi'$ along the c$^{\ast}$-direction is much weaker than that along the a$^{\ast}$-direction,
while $\chi''$ is suppressed strongly as well as that along the a$^{\ast}$-direction is (see Figs. 3 (c) and (d)).
The suppression of $\chi'$ implies a strong nonlinearity of the magnetization and the field dependence of $\chi''$ indicates
that the anomalous slow dynamics in the helimagnetic state can be easily suppressed by magnetic field.
For the quantitation of the nonlinearity of the magnetization, we estimated the nonlinear susceptibility $\chi_2$.
If $\hac$ is much smaller than $H$, $\chi'$ can be treated as the differential magnetization given by 
\begin{eqnarray}
\chi' \simeq \left. \frac{\mathrm{d}M}{\mathrm{d}H}\right|_{H = H_{\mathrm{DC}}} = \chi_0 + 3 \chi_2 H^2 + 5 \chi_4 H^4 + \cdots.
\end{eqnarray}
Then, the nonlinear susceptibility  $\chi_2$ was obtained from the slant of the $\chi'$ vs $H^2$ plots in the limit of $H^2 \rightarrow 0$.
In this analysis, demagnetization correction was operated carefully.
The obtained $\chi_2$ was shown in Fig.\ref{H-dep} (e).
We can find a very strong negative peak of $\chi_2$ at $\TN$ along the a$^*$-direction.

\begin{figure}[h]
\begin{center}
\begin{minipage}{12pc}
\includegraphics[width=12pc]{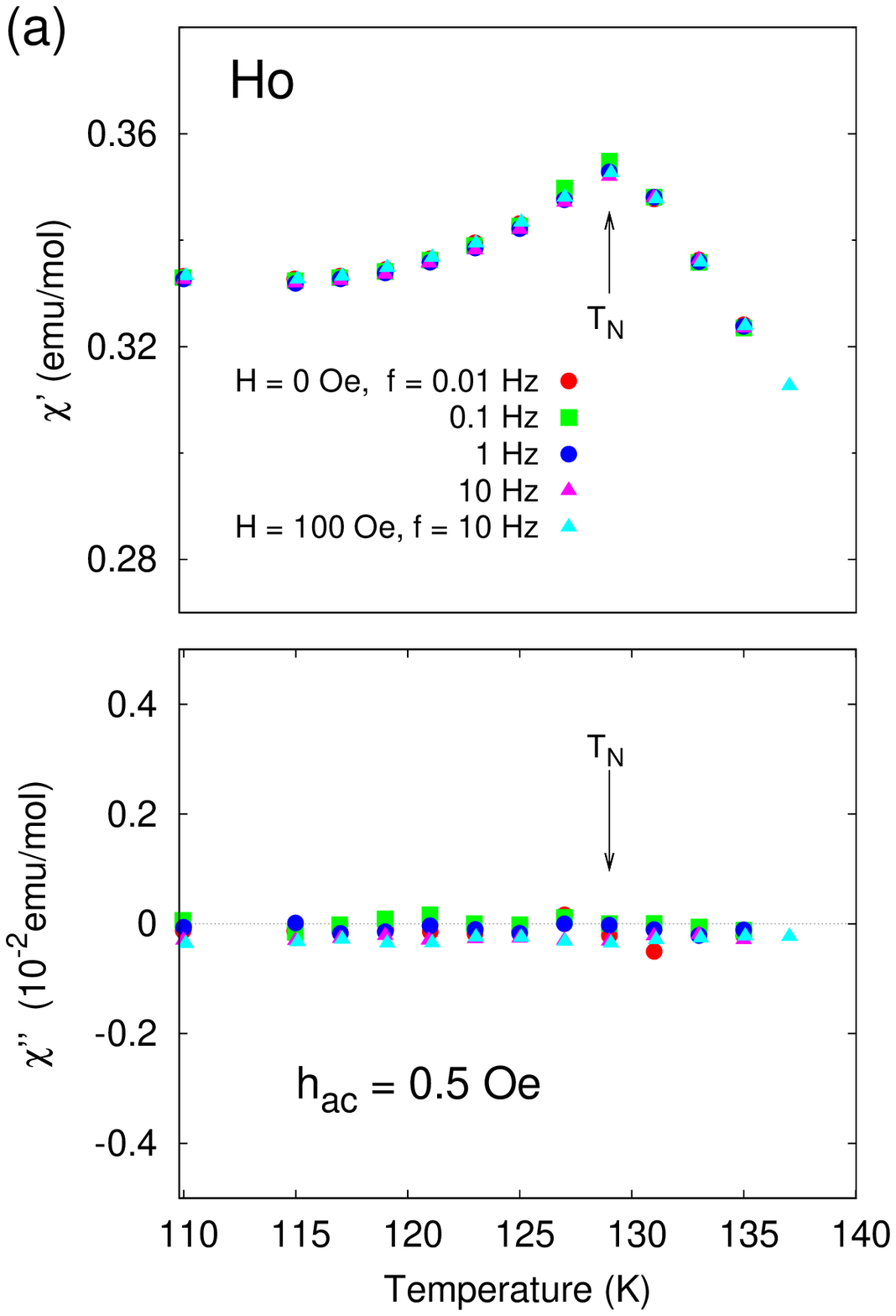}
\end{minipage}\hspace{2pc}%
\begin{minipage}{12pc}
\includegraphics[width=12pc]{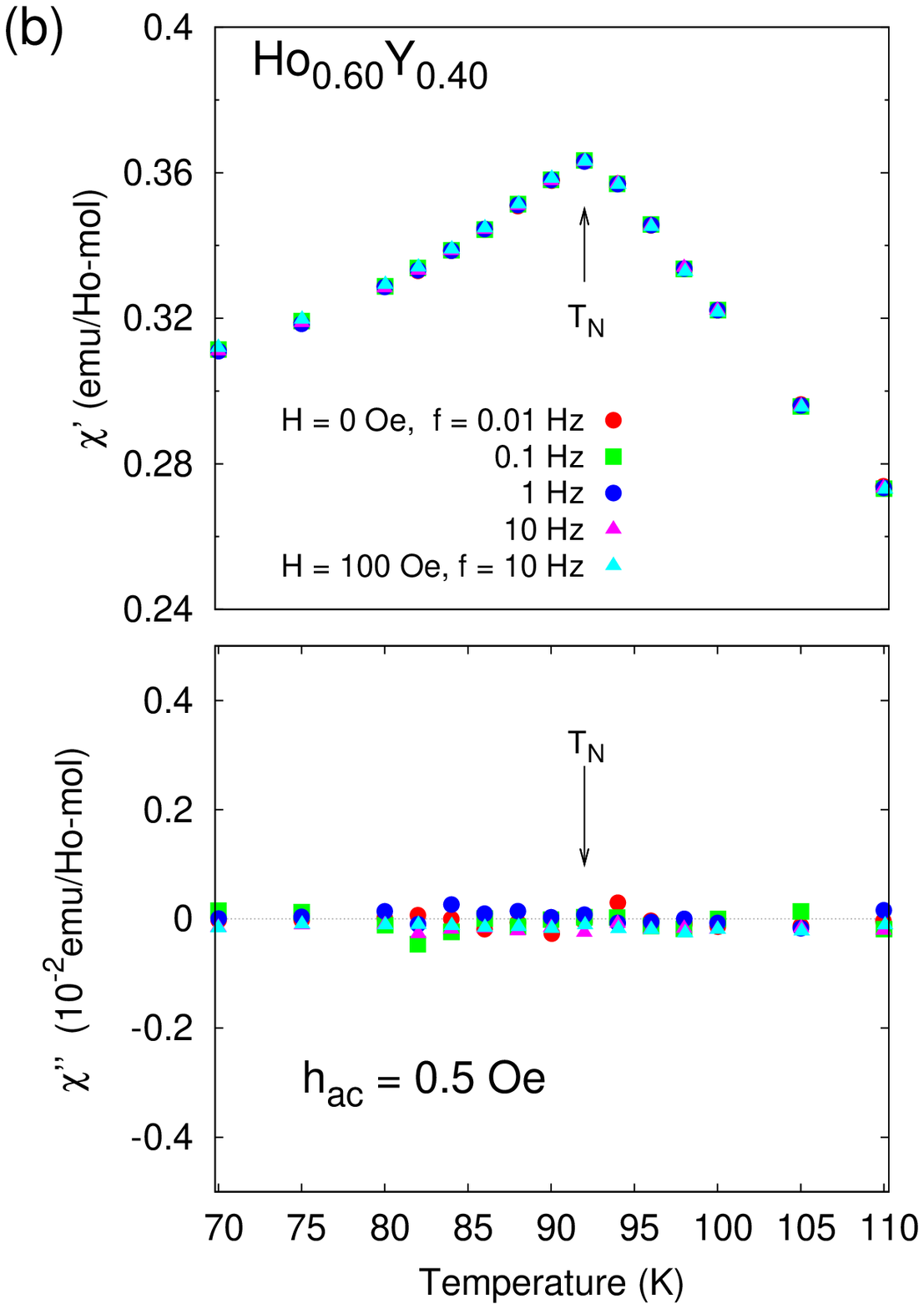}
\end{minipage} 
\caption{\label{HoY}Temperature dependences of the real and imaginary parts of the ac-susceptibility, $\chi'$ and $\chi''$, in (a) Ho and (b) Ho$_{0.60}$Y$_{0.40}$.}
\end{center}
\end{figure}

We performed the same ac-susceptibility measurements in the similar rare-earth helimagnets Ho and Ho$_{0.60}$Y$_{0.40}$ for comparison.
These materials also exhibit para-helimagnetic transition at $\TN$ = 130 K and 92 K.
Figure \ref{HoY} shows temperature dependences of $\chi'$ and $\chi''$ around $\TN$ in polycrystalline Ho and Ho$_{0.60}$Y$_{0.40}$ with the frequency of 0.01 Hz $\leq f \leq$ 10 Hz.
The magnitude of the ac-magnetic-field was 0.5 Oe.
To confirm the nonlinearity of the magnetization, the measurement with applying the dc-magnetic-field of 100 Oe was also performed.
No frequency dependence of $\chi'$ was observed in both materials.
Correspondingly, an increase of $\chi''$ was not observed at all in this frequency range. 
No dc-magnetic-field dependences of $\chi'$ and $\chi''$ were observed in these helimagnets.

Here, we discuss on the peculiar properties newly observed in Gd$_{0.62}$Y$_{0.38}$,
the anomalous slow dynamics and the strong nonlinearity of the magnetization in the helimagnetic state.
In the diluted system such as Gd$_{0.62}$Y$_{0.38}$, randomness can be an origin of the slow dynamics, as the case of spin glasses.
Nevertheless, the increase of $\chi''$ and nonlinearity of the magnetization were not observed in the similar dilute rare-earth helimagnet Ho$_{1-x}$Y$_{x}$ (see Fig. \ref{HoY} (b)).
Therefore, we conclude that the randomness due to the dilution is not the origin of the slow dynamics and nonlinearity of the magnetization in Gd$_{1-x}$Y$_{x}$. 
It should be noted that similar nonlinearity has been reported in Gd$_{1-x}$La$_{x}$ in a previous work \cite{Larica}.
The previous and our experimental results suggest that these anomalous behaviors result from the weak magnetic anisotropy of Gd-moment.
In particular, the in-plane anisotropy is vanishingly weak in Gd$_{1-x}$Y$_{x}$ \cite{Jensen, Brooks}.
And hence, the facts found in this work, the increase of $\chi''$ and nonlinearity of the magnetization observed remarkably only in-plane direction, are consistent with this suggestion.

Now, we propose two different speculations on the origin of the anomalous slow dynamics in Gd$_{1-x}$Y$_{x}$.
The first one is a possible collective motion of a large number of spins in the helimagnetic structure.
A excitation mode that spins rotate dynamically in the helical plane with keeping its helical spin arrangement, which can be excited by the ac-magnetic field along the in-plane direction, has a very low excitation energy
 because of very weak magnetic in-plane anisotropy in Gd$_{1-x}$Y$_x$.
Such a low-energy collective spin excitation can be responsible for the slow dynamics.
In this excitation mode, the spins themselves can be rotated by the ac-magnetic-field, however, the vector chirality is preserved and a motion of the chiral-domains emerges.
It suggests that a chirality is decoupled from spins in the helimagnetic phase of  Gd$_{1-x}$Y$_x$.
In Gd$_{1-x}$Y$_{x}$, the inversion symmetry is preserved globally,
and hence, the left-handed and right-handed chiral domain should coexist in the helimagnetic state.
The domain-wall of these two chiral domain can be also moved by ac-magnetic-field very slowly.
By the dc-magnetic field or the magnetic anisotropy, spins are pinned along the magnetic field
 or the magnetic easy direction, and the chiral-domain and chiral-domain-wall motions can be fixed.
The suppression of $\chi''$ by applying the magnetic field and the lack of $\chi''$ in Ho and Ho$_{1-x}$Y$_{x}$ are manifestations of these pinning effects.
The second is a long-time variation of the period of the magnetic structure, which was recently observed in Ca$_3$Co$_2$O$_6$ \cite{Moyoshi}.
It is known that the period of the helimagnetic structure gradually changes  with changing temperature in zero magnetic field in Gd$_{1-x}$Y$_{x}$.
It indicates that energy levels of states with a little different wave vector $\bm{q}$ are close and degenerate mostly each other, which may result the slow dynamics.
The observation of the time-variation of the magnetic wave vector by the neutron scattering experiments with epithermal neutron or X-ray magnetic scattering is exciting future issue.

\section{Conclusion}
An anomalous slow dynamics and a strong nonlinearity of magnetization were observed in the metallic helimagnet Gd$_{1-x}$Y$_x$.
These peculiar phenomena should result from the weak magnetic anisotropy of Gd-moments.
The anomalous slow dynamics may be chiral-domain and chiral-domain-wall motions or a long-time variation of the period of the magnetic structure.

\ack
This work was partly  supported by a Grant-in-Aid for Scientific Research, the Ministry of Education, Culture, Sports, Science and Technology, Japan (the Global COE Program "Integrated Materials Science" on Kyoto University and Priority Areas "Novel States of Matter Induced by Frustration").

\end{document}